\documentclass[floatfix,twocolumn,10pt,pra,longbibliography,aps]{revtex4-1}

\usepackage{amssymb}
\usepackage{amsmath}
\usepackage{graphicx}
\usepackage[colorlinks, linkcolor=blue, citecolor=blue, urlcolor=blue, breaklinks=true]{hyperref}
\usepackage[english]{babel}

\begin{document}

\title{Many-Body Quantum Spin Dynamics with Monte Carlo Trajectories\\ on a Discrete Phase Space}
\author{J.~Schachenmayer}
\author{A.~Pikovski}
\author{A.~M.~Rey}
\affiliation{JILA, NIST \& Department of Physics, University of Colorado, 440 UCB, Boulder, CO 80309, USA}

\date{\today}

\begin{abstract}
Interacting spin systems are of fundamental relevance in different areas of physics, as well as in quantum information science, and biology. These spin models represent the simplest, yet not fully understood, manifestation of quantum many-body systems. An important outstanding problem is the efficient numerical computation of dynamics  in large spin systems. Here we propose a new semiclassical method to study many-body spin dynamics in generic  spin lattice models.  The method is based on a discrete Monte Carlo sampling in phase-space in the framework of the so-called truncated Wigner approximation. Comparisons  with analytical and numerically exact calculations demonstrate the power of the technique. They show that it correctly reproduces the dynamics of  one- and two-point correlations and spin squeezing at short times, thus capturing entanglement. Our results open the possibility to study the quantum dynamics accessible to recent experiments in regimes where other numerical methods are inapplicable.
\end{abstract}

\maketitle

\section{Introduction}

Controlled experimental observation of
nonequilibrium spin dynamics has recently become possible~\cite{bloch_quantum_2012,blatt_quantum_2012,aspuru-guzik_photonic_2012}.
Large spin systems with long-range interactions have been realized e.g.~with   polar molecules~\cite{yan_observation_2013,hazzard_many-body_2014},  Rydberg atoms~\cite{low_experimental_2012,schaus_observation_2012,schaus_dynamical_2014},  and trapped ions~\cite{islam_emergence_2013,jurcevic_observation_2014,richerme_non-local_2014,Britton:2012}. Key  aspects of quantum dynamics, such as the buildup of long-range correlations, entanglement, and the  propagation of information are insufficiently understood,
partly  due to the absence  of   appropriate tools to  calculate the time evolution in complex quantum systems.

Current techniques are not suitable for the investigation of quantum dynamics in generic large spin systems. For example, numerical time-dependent density matrix renormalization group (t--DMRG) methods~\cite{vidal_efficient_2004,white_real-time_2004,daley_time-dependent_2004} become inefficient in higher dimensional systems; perturbative and Keldysh techniques~\cite{Kamenev2004}, as well as kinetic theories and phase space methods~\cite{blakie_dynamics_2008,polkovnikov_phase_2010,altland_nonadiabaticity_2009}   are limited to weakly interacting,  close-to-equilibrium or short-time situations; cluster expansions~\cite{hazzard_many-body_2014} work only for dilute samples  with  moderately short-ranged interactions.
The development of new  numerical techniques is, therefore, of immediate relevance. In this work, we advance  in this direction by introducing a semiclassical phase-space method  that  we refer to as the discrete truncated Wigner approximation (DTWA). With this relatively easily programmable  method we can calculate nonequilibrium dynamics in systems of thousands of spins and in arbitrary dimensions.

The DTWA is a semiclassical method which is based, in contrast to existing techniques, on the sampling of a discrete Wigner function.
Standard phase-space methods, such as the truncated Wigner approximation (TWA), replace the quantum-mechanical time evolution by a semiclassical evolution via classical trajectories. The quantum uncertainty in the initial state is accounted for by an average over different initial conditions~\cite{blakie_dynamics_2008,polkovnikov_phase_2010} determined by a continuous Wigner function. In contrast, the use of discrete Wigner functions enables us to quantitatively access dynamics in generic spin lattice models, including oscillations and revivals of single particle observables and correlation functions that are not captured by the TWA.

This paper is organized as follows: In Sec.~\ref{sec:twa} the traditional TWA is reviewed, and in Sec.~\ref{sec:dtwa} our DTWA technique is introduced. Sec.~\ref{sec:ram} contains a benchmark of the improvement provided by the DTWA via comparisons of dynamics of single-spin observables, correlation functions and spin squeezing for a model with Ising and XY interactions. Sec.~\ref{sec:concl} provides a conclusion and an outlook.

\section{Method}
\subsection{Semiclassical phase-space sampling}
\label{sec:twa}

\begin{figure*}[t]
		\includegraphics[width=0.8\textwidth]{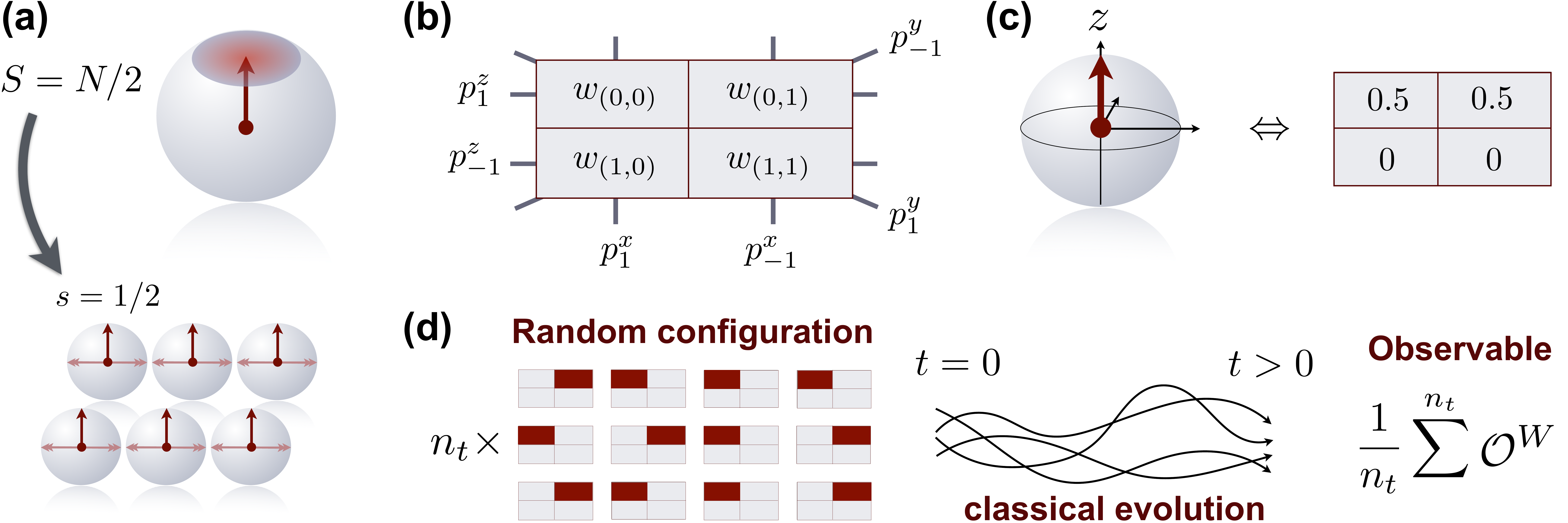}
\caption{{\it Discrete phase space and the DTWA.} (a) Our method considers the quantum uncertainties of $N$ spin-$1/2$ particles individually, rather than the noise of the collective spin $S=N/2$.
(b) The quantum physics of a spin-$1/2$ particle can be fully described by a discrete four-point Wigner quasiprobability distribution, $w_{(p,q)}$. The probability for a spin to point  along the $\pm z$, $\pm y$, and $\pm x$ directions ($p_{\pm 1}^{x,y,z}$) is given by the sum over the vertical, diagonal, and horizontal lines in phase space, respectively \cite{wootters_wigner-function_1987}. (c) Discrete Wigner function of a spin pointing along $z$. (d) The idea behind the DTWA  is to: i) randomly sample the phase points for each spin according to $w_{(p,q)}$; ii) calculate the time evolution according to classical equations; and iii) evaluate observables in phase space from the statistical mixture of $n_t$ trajectories.}
	\label{fig:1}
\end{figure*}

The mapping between the Hilbert space of a quantum system  and its corresponding phase space
(known as Wigner-Weyl transform) can be accomplished through the so-called phase-point operators  $\hat{A}$. In terms of classical phase-space  variables $ {\bf p}$ and $ {\bf q}$ (we set $\hbar=1$ in  this paper), the phase-point operators can be written as  \cite{Fano1957,wootters_wigner-function_1987}: $\langle {\bf q}'|\hat{A}({\bf p},{\bf q})|{\bf q}''\rangle =\frac{1}{(2\pi )^{D}}\delta\big({\bf q}-\frac{{\bf q}'+{\bf q}''}{2}\big)e^{i{\bf p}\cdot({\bf q}'-{\bf q}'')}$ with $D$ the phase-space dimension. They relate the density matrix of the quantum system $\hat{\rho}$ to a  quasiprobability distribution (generally nonpositive) known as the Wigner function $W$~\cite{Fano1957},
which is given by $W({\bf p},{\bf q})={\rm Tr}[\hat{\rho}  \hat{A}({\bf p},{\bf q})]$. Any operator  $\hat{O}({\bf p},{\bf q})$ can be mapped to a function over the classical phase space, the so-called Weyl symbol  $\mathcal{O}_W({\bf p},{\bf q})={\rm Tr}[ \hat{{O}}\hat{A}({\bf p},{\bf q})]$. To compute the time evolution of the expectation value of an operator, its time-evolved Weyl symbol has to be averaged over the phase space with the corresponding Wigner function:
$
\langle \hat{{O}}\rangle(t) = \iint \! d{\bf p}d{\bf q}  \, \mathcal{O}_W({\bf p},{\bf q})W({\bf p},{\bf q};t)
$.
In general, however, it is not possible to compute the time evolution exactly. The TWA~\cite{blakie_dynamics_2008,polkovnikov_phase_2010} approximates the dynamics by taking only first-order quantum fluctuations into account. In the Heisenberg picture the Wigner function is fixed to its initial state $W({\bf p},{\bf q})\to W({\bf p}_0,{\bf q}_0)$ and the Weyl symbol evolves in time. The TWA makes the approximation that the Weyl symbols follow a classical evolution. They are obtained by solving Hamilton's equations of motion for ${\bf p}_{cl}(t)$, ${\bf q}_{cl}(t)$ with the initial conditions $({\bf p}_0, {\bf q}_0)$, and one puts
$\mathcal{O}_W({\bf p},{\bf q})(t) \to \mathcal{O}_W({\bf p}_{cl}(t),{\bf q}_{cl}(t))$; thus:
\begin{align}
\langle \hat{{O}}\rangle(t) \approx \iint \! d{\bf p_0}d{\bf q}_0  \, \mathcal{O}_W({\bf p}_{cl}(t),{\bf q}_{cl}(t))W({\bf p}_0,{\bf q}_0)
.\label{eq:cont_twa}
\end{align}

Generalizations of this continuous formulation  to     $N$ spin-$1/2$ particles  have been developed, e.g.~by  means of a  spin-coherent state representation,   valid up to $1/N$ corrections. Generically, the Wigner function is approximated by a smooth positive Gaussian-like distribution in the collective spin variables. For example if all spins are  pointing along the $z$ axis, the Wigner function can be approximately written  as~\cite{altland_nonadiabaticity_2009} $W(S_\perp, S_z)\approx 1/(\pi S) e^{-S_\perp^2/S}\delta(S_z-S)$, with $S=N/2$, $S_\perp=(S_x^2+S_y^2)^{1/2}$, and $S_z$  the transverse and longitudinal spin components of the collective classical spin, respectively. This Wigner function has a clear  interpretation: If each  spin initially points along the $z$ direction, the transverse spin components must fluctuate as $\langle S_\perp^2\rangle \sim S$ due to the Heisenberg  uncertainty principle.

This Gaussian  TWA generally captures aspects of the quantum spin dynamics at short times but  lacks important ones such as the  revivals [see e.g.\ Fig.~\ref{fig:2}(a)],  ubiquitous in   discrete quantum systems, and it is mainly limited  to dealing with  the collective   dynamics \cite{polkovnikov_phase_2010}. For interactions with spatial structure the dynamics quickly takes  the system out of the collective-spin Hilbert space and the TWA breaks down.  In the continuous phase-space picture,  ways to overcome these shortcomings have been proposed using hidden variables \cite{Davidson2014} or more complex representations of the Wigner function \cite{Cohen1986,Takahashi1975,polkovnikov_phase_2010}.
Here we propose and test a different approach, which uses discrete phase spaces for each individual spin in conjunction with a Monte Carlo
sampling.

\subsection{Discrete phase-space sampling}
\label{sec:dtwa}

For systems with discrete degrees of freedom, it is possible to define a ``discrete phase space'' in many ways (see Ref.~\cite{wootters_wigner-function_1987} and references therein). We use four phase points to describe a single spin-1/2, which we denote as $\alpha \equiv(q, p) \in \{(0,0),(0,1),(1,0),(1,1)\}$ as introduced by Wootters~\cite{wootters_wigner-function_1987,wootters_picturing_2003}. For each phase-space point one can define a phase-point operator $\hat A_{\alpha}$, the Weyl symbols $\mathcal{O}_\alpha^{W}={\rm tr}(\hat {{O}} \hat A_{\alpha})/2$ (or  inversely $\hat {{O}}=\sum_\alpha \hat A_{\alpha} \mathcal{O}_\alpha^{W}$), and  the Wigner function $w_\alpha$, which is the Weyl symbol for the density matrix.  A  set of phase-point operators is given by \cite{wootters_wigner-function_1987}
 \begin{align}
 \hat A_{\alpha}= \hat {\wp} ({\bf r}_\alpha),  \qquad {\hat \wp}({\bf r})\equiv (1+ {\bf r}\cdot\hat{ {\boldsymbol \sigma}})/2
 \end{align}
 with ${\bf r}_{(0,0)}= (1,1,1)$,  ${\bf r}_{(0,1)}= (-1,-1,1)$,  ${\bf r}_{(1,0)}=  (1,-1,-1)$, and  ${\bf r}_{(1,1)}=  (-1,1,-1)$. Here  $\hat{ {\boldsymbol \sigma}}=(\hat{\sigma}^{x},\hat{\sigma}^{y} ,\hat{\sigma}^{z} )$ are the Pauli matrices. In a  many-body system with $N$ spin-$1/2$ particles the discrete phase space has $4^N$ points we denote as ${\boldsymbol \alpha}=\{\alpha_1,\alpha_2,\dots \alpha_N\}$. Analogously to Eq.~\eqref{eq:cont_twa},
 we can now formulate the DTWA  on this discrete phase space as
\begin{align}
	\langle \hat O \rangle (t) =\sum_{\boldsymbol \alpha} w^{}_{\boldsymbol \alpha}(0) \mathcal{O}^W_{\boldsymbol \alpha} (t)
	\approx \sum_{\boldsymbol \alpha} w^{}_{\boldsymbol \alpha}(0) \mathcal{O}^{W,{\rm cl.}}_{\boldsymbol \alpha} (t),
	\label{eq:dtwa}
\end{align}
where $w^{}_{\boldsymbol \alpha}(0)$ is the initial Wigner function on the  discrete many-body phase space and $\mathcal{O}^{W,{\rm cl.}}_{\boldsymbol \alpha} (t)$ is the classically evolved Weyl symbol (see Appendix~\ref{app-class} for more on the classical equations of motion). Numerically, we can solve Eq.~\eqref{eq:dtwa} by statistically  choosing [according to  $w^{}_{\boldsymbol \alpha}(0)$] a large number $n_t$ of random initial spin configurations. Each ``trajectory'' is evolved in time according to the classical equations of motion and the expectation value in Eq.~\eqref{eq:dtwa} is estimated via statistical averaging (error $\sim1/\sqrt{n_t}$). We find that the required $n_t$ does not depend on the system size, but rather on the observable under consideration (see also Appendix~\ref{app-conv}).

 As an example of how to construct the initial Wigner function, we again consider an initial state with all spins pointing along the $z$ direction.  The initial density matrix factorizes $\hat \rho(0)=\prod_{i=1}^N {\hat \wp}^{[i]}(\hat{z})$ (the superscript ${[i]}$ denotes the Hilbert or phase space for spin $i$), and thus,  $w^{}_{\boldsymbol \alpha}(0)=\Pi_{i=1}^N w^{[i]}_{\alpha_i}$. Here, $w^{[i]}_{\alpha_i} ={\rm Tr}[{\hat \wp}^{[i]}(\hat{z}) \hat A_{\alpha_i}]/2$ is given by $w^{[i]}_{(0,0)}=w^{[i]}_{(0,1)}=1/2$, and $w^{[i]}_{(1,0)}=w^{[i]}_{(1,1)}=0$ for every spin $i$ (cf.~Fig.~\ref{fig:1} for an illustration). Note that for this initial state,  all quasiprobabilities are positive, which is important for the numerical sampling. The sum along each ``phase-space line'' of the discrete Wigner function can be associated with the probability of a measurement outcome \cite{wootters_wigner-function_1987}, similarly to the continuous variable case. As shown in Fig.~\ref{fig:1}, here $w^{[i]}_{(0,0)}=w^{[i]}_{(0,1)}=1/2$ means that the probability to find an individual spin pointing along the $+z$-direction is $1$ (sum over the first row), while the probabilities to find it along $+x$ or $-x$ (sum over each column) are $50\%$ each. Equally, the probabilities to find it along $+y$ or $-y$ are 50\% each (sum over each of the two diagonals). Note that  this discrete sampling properly  accounts for the quantum-mechanical probability distribution of the $x, y,z$ spin components of a qubit in the sense that all moments are reproduced correctly: $\langle(\hat{\sigma}^{x,y})^k \rangle=(1+(-1)^k)/2$, $\langle(\hat{\sigma}^{z})^k \rangle=1$, with $k$ a positive integer.

\section{Dynamics using the DTWA}
\label{sec:ram}

\begin{figure*}[t]
	\includegraphics[width=1\textwidth]{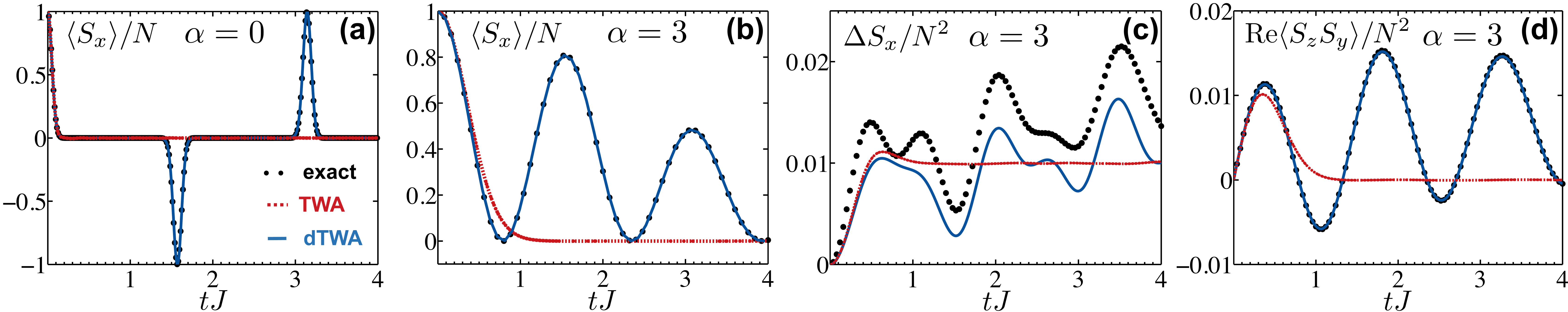}
	\caption{{\it Dynamics for Ising interactions.} Circles denote the exact solution, dashed lines are traditional TWA results, solid lines denote  DTWA results, for 1D, $N=100$ spins. (a-b) Evolution of $\langle S_x \rangle$, for all-to-all (decay exponent $\alpha=0$) and dipolar ($\alpha=3$) interactions, respectively. Traditional TWA captures only the initial decay and no oscillations or revivals. In contrast, DTWA becomes exact (on top of the black symbols). (c/d) The evolution of the correlation functions $\Delta S_x = \langle S_x^2 \rangle -\langle S_x \rangle^2$ and $\text{Re} \langle S_y S_z \rangle$ for dipolar interactions. While the latter one is exactly captured in DTWA, $\Delta S_x$ shows deviations. DTWA improves traditional TWA predictions in all panels.}
	\label{fig:2}
\end{figure*}

To demonstrate the accuracy of the DTWA we consider a system of $N$ two-level systems arranged on a lattice with $M$ sites with  dynamics  governed by the Hamiltonian
\begin{align}\label{H}
\hat{H}=\frac{1}{2} \sum_{i \neq j} \left[ \frac{J^\perp_{ij}}{2}  ( \hat{\sigma}_i^x \hat{\sigma}_j^x + \hat{\sigma}_i^y \hat{\sigma}_j^y)   +  J^z_{ij} \hat{\sigma}_i^z \hat{\sigma}_j^z  \right] + \Omega \sum_i \hat{\sigma}_j^x
\end{align}
We consider an initial  product state in which all spins are aligned along the $x$ axis. The interactions are assumed to decay as a function of the distance with a decay exponent $\alpha$, and are allowed to be spatially inhomogeneous: e.g.~$J_{ij}^{\perp/z}\equiv J[1-3\cos(\theta)^2]/|{\bf r}_{ij}|^\alpha$. Here, ${\bf r}_{ij}$ is the vector connecting spins on sites  $i$ and $j$ and $\theta$ is the angle it makes with the quantization axis (chosen along $z$). We  discuss two specific  cases, Ising ($J^\perp_{ij}=0$) and XY ($J^z_{ij}=0$) interactions. In addition to the interactions we allow for a transverse field with strength $\Omega$. The Ising limit is naturally realized in  experiments with ion traps (both in  1D  \cite{jurcevic_observation_2014,richerme_non-local_2014} or 2D \cite{Britton:2012} geometries) and Rydberg atoms in 2D \cite{schaus_dynamical_2014}; the XY limit dynamics have been realized  in polar  molecules in optical lattices \cite{yan_observation_2013,hazzard_many-body_2014},  magnetic atoms \cite{de_paz_nonequilibrium_2013}, F{\"o}rster resonances in Rydberg atoms \cite{nipper_atomic_2012}, and  as an effective Hamiltonian in  trapped ions  with a superimposed  large  transverse field \cite{richerme_non-local_2014}.

The classical equations of motion  (for classical spin components $s_i^{x,y,x}$)  corresponding to Hamiltonian~\eqref{H} are given in Appendix~\ref{app-class}. The  DTWA method simply consists of a numerical integration of the classical  equations of motion for many different random initial conditions. While for each site $i$, the initial condition of the classical spin component along $x$, $s^x_i=1$ is fixed, the initial conditions for the spin components in the orthogonal directions are randomly chosen as $s^y_i,s^z_i=\pm1$ (as motivated in Sec.~\ref{sec:dtwa}). Final expectation values of observables are calculated by averaging the results for the corresponding observable over all initial conditions, i.e.~all trajectories.

\subsection{Ising interactions}

\begin{figure*}[t]
	\includegraphics[width=0.8\textwidth]{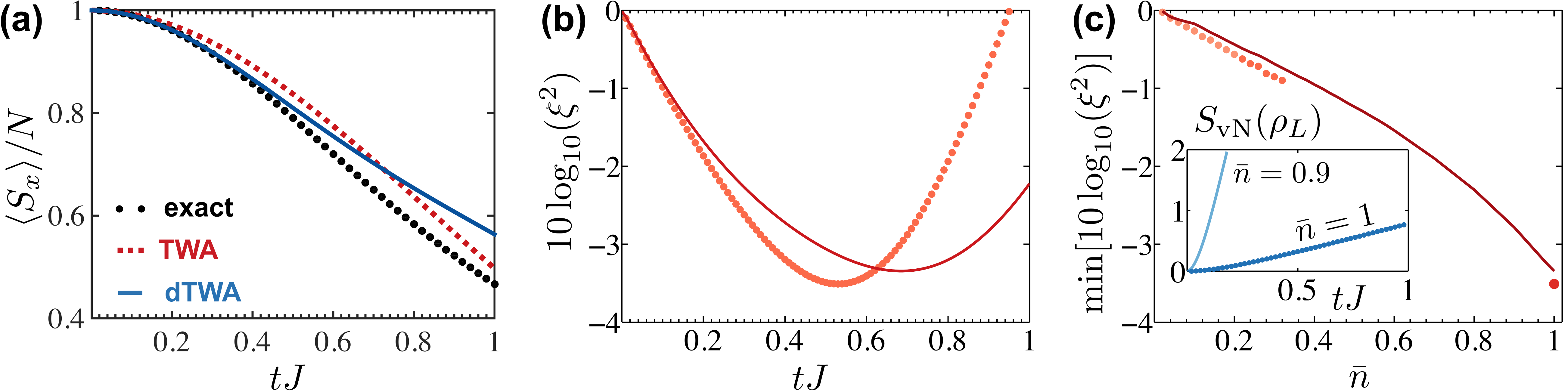}
	\caption{{\it Dynamics for XY interactions.} Circles denote exact, solid lines DTWA, and dotted lines traditional TWA results. (a) Time evolution of $\langle S_x \rangle$  for the XY model in 1D, for $N=100$ spins and with long-range interactions with decay exponent $\alpha=3$. In contrast to the Ising case DTWA is only exact for short times only but can capture longer times than traditional TWA. (b) Time evolution of the spin-squeezing parameter $\xi$. The DTWA gives exact results for short times and good estimates for the achievable $\xi$. (b) Achievable $\xi$ as a function of the filling fraction (averaged over $1000$ random configurations). Exact diagonalization (ED) for $N<20$, t-DMRG otherwise. Inset: Rapid increase of the entanglement entropy  for noninteger filling $\bar n<1$ (single configuration), rendering t-DMRG calculations inefficient in this regime.}
	\label{fig:3}
\end{figure*}

We now consider Ising interactions, $J^\perp_{ij}=0$, $\Omega=0$, in Eq. \eqref{H}.
In this limit, exact analytical expressions for the dynamics exist \cite{Lowe-Norberg-1957,foss-feig_nonequilibrium_2013,worm_relaxation_2013} and can be used to benchmark the DTWA. The dynamics of observables involving the collective spin   ${\boldsymbol S}=(\langle S_x \rangle ,\langle S_y \rangle ,\langle S_z\rangle)^T \equiv \sum_n \langle \hat {\boldsymbol \sigma}_n \rangle$  for a system with $N=100$ spins in a one-dimensional chain with $M=100$ sites  (oriented along $x$)  are shown in Fig.~\ref{fig:2}. We calculate the time evolution of  $S_x$
as well as the collective correlation functions $\Delta S_x = \langle S_x^2 \rangle -\langle S_x \rangle^2$ and ${\rm Re}\langle S_y S_z \rangle$.
We consider  the case of all-to-all (decay exponent $\alpha=0$) and short-ranged dipolar interactions ($\alpha=3$). In the all-to-all case $\langle S_x \rangle$ shows revivals at times that are multiples of $\pi/2J$. In contrast to the  traditional Gaussian  TWA,  which captures only the initial decay of $\langle S_x \rangle$ and misses the  revivals, the DTWA fully  reproduces the exact dynamics (a similar effect has been seen in Ref.~\cite{hush_number-phase_2010}). The case $\alpha=3$   exhibits an oscillatory  dynamics, perfectly accounted for by the  DTWA solution, but not captured by the traditional  TWA. The DTWA calculations for ${\rm Re}\langle S_y S_z \rangle$ also  show  perfect agreement. Deviations are visible for the correlation $\Delta S_x$, however, the oscillatory dynamics  is still better reproduced in the DTWA than in the traditional TWA.

We understand the  agreement between the Ising solution and the DTWA analytically. For a particular spin $n$,  the exact solution for the time evolution is $\langle \hat \sigma_{n}^x \rangle(t)=\prod_{i\neq n}^N \cos(2tJ^z_{in})=\sum_{\bf m} \cos[2t\sum_{a=1}^{N}(J^z_{na} m_a )]/2^N$, where ${\bf m}=\{ m_1,m_2,\dots,m_N\}$ and each of the the $m_a$ takes the values $\pm 1$. The classical equation of motion for the $x$ component of spin is $s_n^x (t)= \cos(2t \beta_n)$ (see Appendix~\ref{app-class}), where $\beta_n=\sum_{i\ne n}^{N} J_{in} s_i^z(t=0)$. Since we initially sample the $z$ component of the spin to be randomly distributed as  $s_i^z(t=0)=\pm 1$,  we can identify $s_i^z(t=0)$ with the parameters $m_i$ in the exact solution and find that the DTWA  becomes equivalent to the exact solution for large $n_t$ and arbitrary numbers of spins $N$. Note that the  traditional TWA approach is valid only in the large-$N$  limit. The comparisons with the exact solution, extended to correlation functions, confirm the excellent agreement of  ${\rm Re}\langle S_y S_z \rangle$ and show the origin of the discrepancies in other two-point correlations (see Appendix~\ref{app-corr} and Fig.~\ref{fig:2}).

\subsection{XY model}

Next we consider the dynamics for the XY model, $J^z_{ij}=0$, $\Omega=0$,  in Eq. \eqref{H}.
Here we study systems with varying filling fractions $\bar n=N/M$, a situation relevant for recent polar molecule experiments \cite{yan_observation_2013}. We focus our attention on the time evolution of spin squeezing, which is a signature of  quantum correlations and  entanglement~\cite{sorensen_many-particle_2001} and is a  resource for enhanced sensitivity in quantum metrology~\cite{giovannetti_advances_2011}. The spin-squeezing parameter is $\xi \equiv\sqrt{N} \min_{{\bf n}_\perp}(\Delta S_\perp)/|{\bf S}|$ where ${\bf S}$ is the total collective spin, $S_\perp = {\bf S} \cdot \bf{n}_\perp$, and the minimum is taken over all unit vectors $\bf {n}_\perp$ (directions) perpendicular to the vector ${\bf S}$. On the Bloch sphere a  squeezed state with $\xi<1$ shows an  elliptical profile of the spin noise distribution~\cite{wineland_spin_1992,kitagawa_squeezed_1993}.

For the XY model no exact solution exists for generic spin systems; however in 1D we can use t-DMRG to calculate the exact dynamics at short times. This technique works as long as the bipartite entanglement in the system, quantified by the half-chain von Neumann entropy $S_{\rm vN}(\rho_L)=-{\rm tr}(\rho_L \log_2 \rho_L)$ (where $\rho_L$ is the reduced density matrix of half the spin chain), remains small \cite{amico_entanglement_2008,eisert_colloquium:_2010,daley_measuring_2012,schachenmayer_entanglement_2013}.  Surprisingly, we find that for intermediate filling fractions, due to the inhomogeneity in $J^\perp_{ij}$, the entropy $S_{\rm vN}$ grows much more rapidly than for $\bar n = 1$ [see inset in Fig.~\ref{fig:3}(b)]. Thus, with reasonable computational resources, exact results could be calculated for only $N\leq 32$ spins and $N=100$ spins in a system with $M=100$ sites.

Our results for a 1D chain of spins along the $x$ direction are shown in Fig.~\ref{fig:3}. In Fig.~\ref{fig:3}(a), as in the Ising case, first we analyze the evolution of $\langle S_x\rangle$. We compare DTWA and TWA results to an exact t-DMRG calculation.  Because of the more complicated XY interactions (the Hamiltonian contains noncommuting terms in contrast to the Ising case), we find that DTWA no longer provides an exact solution. Still, it can capture the evolution on the short time scale $tJ \lesssim 0.4$, and significantly improves the traditional TWA result, which, in this example, captures only times $tJ \lesssim 0.1$.

\begin{figure*}[t]
		\includegraphics[width=1\textwidth]{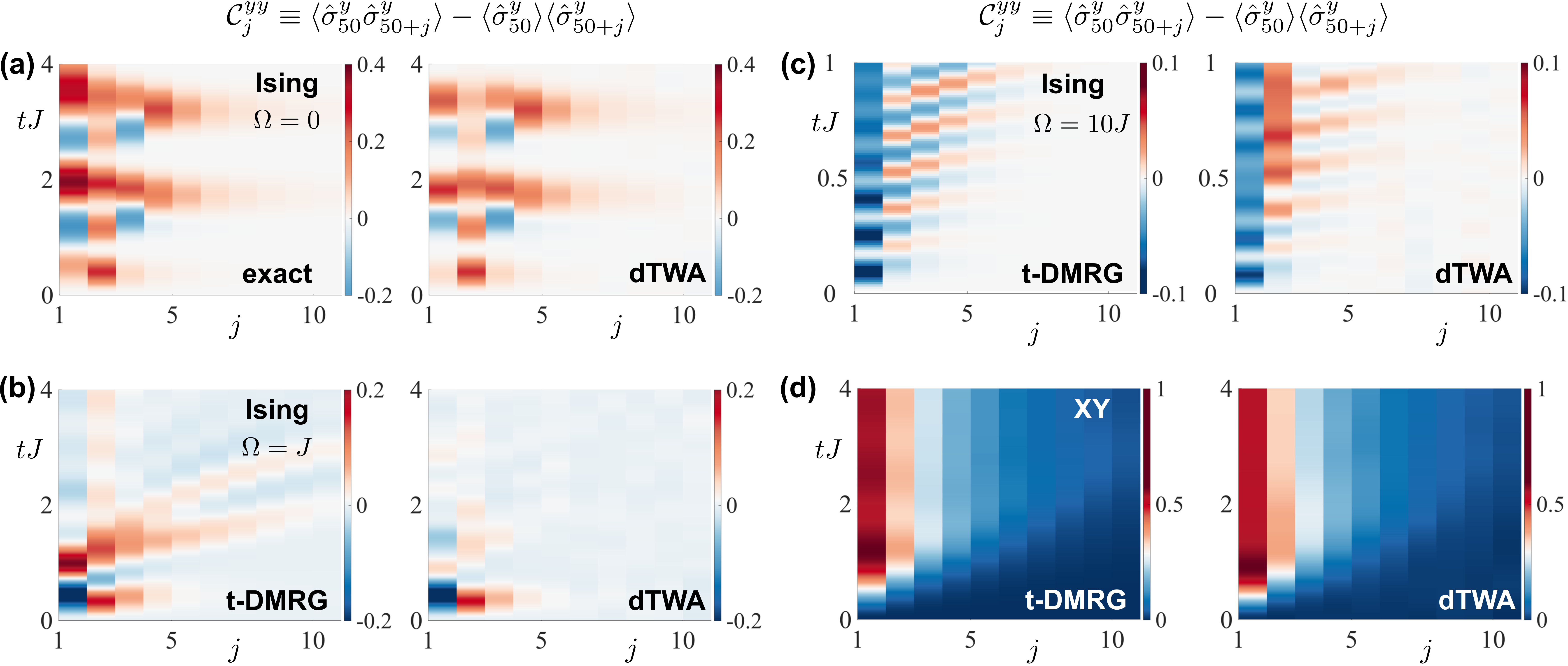}
	\caption{{\it Spreading of correlations.} Time evolution of $\mathcal{C}^{yy}_{j}$ in a system with $N=100$ spins on $M=100$ sites for long-range interactions with decay exponent $\alpha=3$. We consider three models [Eq.~\eqref{H}]: (a) Ising interactions, (b),(c) Ising interactions plus a transverse field term $\Omega \sum_i \hat \sigma_i^x$, and (d) XY interactions. We compare exact or t-DMRG results (left-hand side) with results obtained from the DTWA (right-hand side).}
	\label{fig:3b}
\end{figure*}

In Fig.~\ref{fig:3}(b) we show the evolution of the spin-squeezing parameter $\xi$ for $\bar n=1$.
We also find here that DTWA agrees on short times and captures almost all of the spin squeezing that is created in the time evolution. In Fig.~\ref{fig:3}(b), we plot the maximally achievable spin squeezing as function of the filling fraction. The DTWA interpolates over the whole range of filling fractions connecting the exactly tractable limits. It is interesting to note that while squeezing implies entanglement (nonseparability)~\cite{sorensen_many-particle_2001}, the type of squeezing we consider here is apparently independent of bipartite entanglement, in the sense that for smaller $S_{\rm vN}(\rho_L)$ we can have large squeezing and vice versa. We note that the connection between entanglement or spin-squeezing and discrete phase spaces has been also explored in Ref.~\cite{marchiolli_spin_2013}.

\subsection{Spreading of correlations \& transverse field}

\begin{figure*}[t]
		\includegraphics[width=0.8\textwidth]{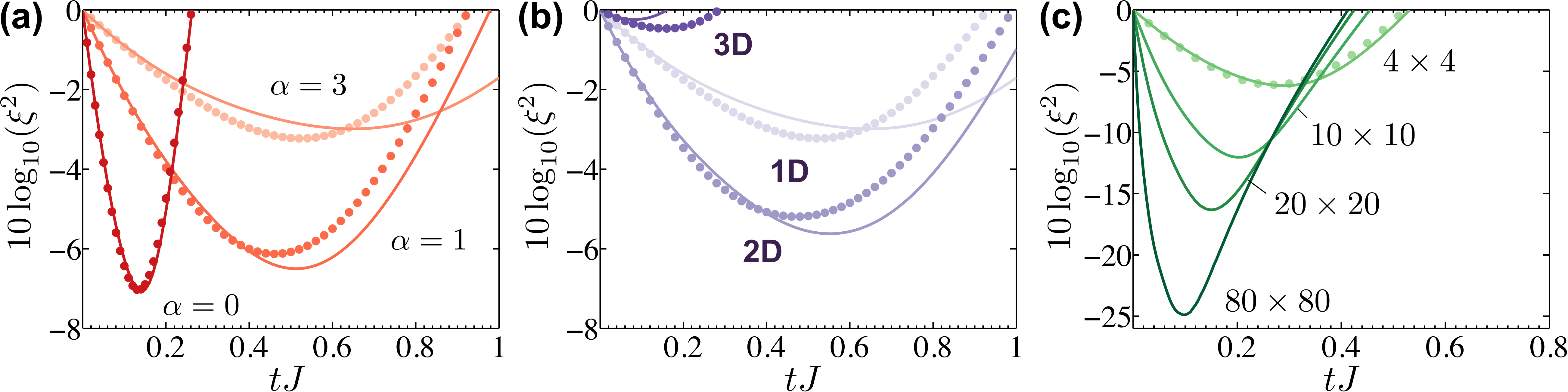}
	\caption{{\it Range of interactions and higher dimensions (XY model).} Circles denote exact diagonalization, solid lines  DTWA results. The time evolution of the spin-squeezing parameter $\xi$  is shown: (a) in 1D ($N=20$ spins) for different decay exponents $\alpha$, and (b) with fixed $\alpha=3$ and for different dimensions (1D: $20 \!\times\! 1 \!\times\! 1$ lattice, 2D: $5 \!\times\! 4 \!\times\! 1$, 3D: $3 \!\times\! 3 \!\times\! 2$). With increasing range of interactions, the DTWA improves, and spin squeezing increases. (c) DTWA prediction for increasingly large 2D systems ($ N \le 6400$) with $\alpha=1$. The exact diagonalization result is shown for the $4 \!\times\! 4$ lattice.}
	\label{fig:4}
\end{figure*}

In order to understand more systematically what types of correlations can be captured by the DTWA, it is instructive to look at the time evolution of spatial correlations in the system. In Fig.~\ref{fig:3b} we analyze the time evolution of $\mathcal{C}^{yy}_{j} \equiv \langle \hat \sigma_{50}^y \hat \sigma_{50+j}^y \rangle- \langle \hat \sigma_{50}^y\rangle \langle \hat \sigma_{50+j}^y \rangle$, i.e.~a spatial connected two-point correlation calculated from the center of a system with $N=100$ spins on $M=100$ sites. Again we consider initially a state with all spins pointing along the $x$ direction. This leads to the fact that initially $\mathcal{C}^{yy}_{j}=0$ for $j\neq 0$.
We study how correlations propagate throughout the system \cite{schachenmayer_entanglement_2013,eisert_breakdown_2013,gong_persistence_2014} in the presence of long-range interactions (decay exponent $\alpha=3$). Besides the Ising and the XY model, we also treat the case of an Ising interaction with a transverse field (we consider a small $\Omega/J=1$ and a large $\Omega/J=10$ field). The addition of a transverse field to the exactly solvable Ising model adds a noncommuting term in the Hamiltonian
and thus enhances  the development of quantum correlations during the dynamics. In this situation  one has to resort to t-DMRG techniques in order to solve for the full quantum dynamics.

As shown in Fig.~\ref{fig:3b}(a),  the dynamics in the pure Ising case  is dominated by oscillations. Here, DTWA can capture long-range correlation dynamics ($j\gg1$) very well, but the amplitudes and frequencies of e.g.~nearest-neighbor correlations ($j=1$) are not captured correctly. This can be understood via the error analysis performed in appendix~\ref{app-corr} [cf.~Eq.~\eqref{eq:corr_error}]. When adding a transversal field of order $J$ [Fig.~\ref{fig:3b}(b)] the agreement becomes  worse. In this case, the total magnetization $S_z$ is no longer  a conserved quantity and  the DTWA fails to capture the propagation of correlations observed in the exact solution. However, as $\Omega$ is increased,  the agreement improves  and the spreading of correlations starts to show up also in the DTWA solution. Note that in the presence of a large transverse field, the conservation of the total  magnetization (now along the transverse field direction) is restored since transitions induced by the Ising term to states that do not preserve it become off resonant. In this large transverse field limit the dynamics  can be mapped back to the dynamics in an effective XY model \cite{richerme_non-local_2014} and although the  DTWA cannot correctly reproduce the oscillation of short-range correlations, it is clearly capable of perfectly reproducing the spreading of  $\mathcal{C}^{yy}_{j}$ correlations through the system with time.

The frequencies of the oscillations observed in Fig.~\ref{fig:3b}(a)-(c) can be understood. Without a transverse field, the exact solution [cf.~Eq.~\eqref{corr-exact}] shows that the correlation functions oscillate with frequencies proportional to the sum of the couplings to the other spins. In the case $\alpha=3$ (short range) the oscillations are dominated by nearest neighbor couplings. In the case of a strong transverse field $\Omega \gg J$, the spins precess rapidly along the external field and thus the correlation functions oscillates roughly with that rate, $\Omega$. The case $\Omega\sim J$ is most complex  since it is close to the critical regime for the  ground state. Nevertheless, in this regime  the correlations  oscillate at a rate $\Omega$, the only energy scale in this case. An interesting question to ask is how dynamics is connected to the ground-state phase transition in this model \cite{rey_preparation_2007,schachenmayer_entanglement_2013}. The DTWA could in the future be used for studies in this direction.

The failure of the DTWA to capture  the spreading of correlations  in generic situations arises from the fact  that
it formally  corresponds to the lowest-order approximation in $\hbar$.
Higher-order quantum corrections that induce dynamics in the Wigner distribution can be incorporated via additional ``quantum jumps'' \cite{polkovnikov_phase_2010}. Although this is out of the scope of the current work, it is a direction that could lead to further improvement of the DTWA method and should be subjected to further investigation.

\subsection{Dependence on range of interactions and dimensionality}

We now check the dependence of the validity of the DTWA on the range of interactions and the dimensionality of the system, focusing on the  XY model. The results are summarized in Fig.~\ref{fig:4}. In Fig.~\ref{fig:4}(a) we analyze the time evolution of the spin-squeezing parameter $\xi$ for a $N=20$ spin system in 1D ($M=20$ sites) for various decay exponents $\alpha$ and for fixed $\alpha$ and different dimensions. As expected we find that longer-ranged interactions lead lo larger amounts of spin squeezing. Remarkably, for longer-ranged interactions the increased squeezing is better captured within the DTWA (also when analyzing the relative error). This can be understood in the limit of all-to-all interactions ($\alpha=0$), where due to the conservation of the total spin the XY model becomes equivalent to the Ising model. In that case, we find (Fig.~\ref{fig:2}) that the DTWA is almost exact. For higher-dimensional systems, we find that the agreement becomes generally better. However due to the inhomogeneity of the coupling constants, the 3D case is difficult to access since the spin squeezing becomes very small. In Fig.~\ref{fig:4}(b) we find that for a small 2D system with $\alpha=1$, the agreement is excellent.

\section{Conclusion and Outlook}
\label{sec:concl}

The discrete truncated Wigner approximation (DTWA) opens the possibility of computing the dynamics of large spin systems in regimes where currently there is no other theoretical tool at hand [see in particular Fig.~\ref{fig:4}(b)]. This is possible since the computational time for a solution of the mean-field equations only scales polynomially in time. Furthermore, the Monte Carlo sampling can be perfectly parallelized, and the number of required samples for statistical convergence depends only on the observable. Here, this enables us to make quantitative predictions for short-time dynamics in systems with up to $6400$ spins in two dimensions, a regime that is, for example, clearly inaccessible to t-DMRG methods.

The results show that the DTWA is able to capture the buildup of spin squeezing (an entanglement witness~\cite{sorensen_many-particle_2001}). On the other hand,
from Eq.~\eqref{eq:dtwa} and the classical equations of motion (see Appendix~\ref{app-class}) it follows that $\langle \hat O \rangle (t) \approx \sum_{\boldsymbol \alpha}^{n_t}  w^{}_{\boldsymbol \alpha}(0)\,{\rm Tr}\left[ \hat O\,\bigotimes_{i=1}^{N} \hat A^{[i]}_{\alpha_i}(t) \right]$, which resembles the time evolution of an expectation value computed from a separable density matrix.  This is not a contradiction, because the phase-point operators $\hat A_{\alpha} $ are, in fact,  not   density matrices. While ${\rm Tr}(\hat A_{\alpha} )=1$, it is possible to satisfy  ${\rm Tr}(\hat A_{\alpha} ^2)>1$; these conditions might be interpreted as a statistical mixture with negative probabilities.

The results we present here demonstrate that the DTWA can be relevant for computing the nonequilibrium dynamics in a variety of recent experimental setups including polar molecules~\cite{yan_observation_2013,hazzard_many-body_2014},  Rydberg atoms~\cite{low_experimental_2012,schaus_observation_2012,schaus_dynamical_2014},  trapped ions~\cite{islam_emergence_2013,jurcevic_observation_2014,richerme_non-local_2014,Britton:2012}, alkaline earth atoms ~\cite{martin_quantum_2013},  and  solid-state systems, such as nitrogen-vacancy centers \cite{Doherty20131,Dolde2013,weber2010}, plasmonic lattices~\cite{plasmonic}, and  photonic crystals~\cite{photonic}.

It would be interesting to extend the DTWA
to deal with open quantum systems  and  to solve for equilibrium states by employing an evolution in imaginary time .
Another direction is to try to combine  higher-order corrections to the TWA \cite{polkovnikov_phase_2010,Davidson2014} with the idea of discrete phase spaces; this could lead to an even more powerful method capable of capturing quantum many-body dynamics for longer time scales.

\section*{Acknowledgements}
We appreciate useful discussions
with  K.~R.~A.~Hazzard, A.~Polkovnikov, and B.~Zhu.  This
work has been financially supported by  JILA-NSF-PFC-1125844,
NSF-PIF-1211914, ARO,  AFOSR, AFOSR-MURI. Computations utilized the Janus supercomputer, supported by NSF (CNS-0821794), NCAR and CU Boulder.

\appendix

\section{Classical equations of motion}
\label{app-class}

To apply the truncated Wigner approximation we have to compute the classical equations of motion for the spin components of each spin $i$: $s^x_i,s^y_i, s^z_i$. In the usual phase-space representation of quantum dynamics, for individual spins these can be obtained from the classical Hamiltonian function:
\begin{align}
{H_C}=\frac{1}{2} \sum_{i \neq j} \left[ \frac{J^\perp_{ij}}{2}  ( s_i^x s_j^x + s_i^y s_j^y)   +  J^z_{ij} s_i^z s_j^z  \right] + \Omega \sum_i s_j^x
\end{align}
via
\begin{align}
\dot{s}^\alpha_i=\{s^\alpha_i,H_C\}= 2 \sum_\beta \epsilon_{\alpha \beta \gamma} s^\gamma_i\frac{\partial H_C}{\partial s^\beta_i},
\end{align}
with $\{.,.\}$ denoting the Poisson bracket and $\epsilon$  the fully antisymmetric tensor.

Alternatively,  the same equations of motion can be obtained via a product ansatz for the phase-point operators. The exact quantum evolution of any observable $\hat O$ is given by $\hat O(t)=\hat U^\dag \hat O(0) \hat U$, where $\hat U=\exp(-{\rm i}t \hat H)$ with $\hat H$ the Hamiltonian of the system. The time evolution of a Weyl symbol on the discrete phase space can thus be written as $\mathcal{O}_\alpha^W(t)={\rm tr} [\hat O(t) \mathcal{\hat A}_{\boldsymbol \alpha}]/2 = {\rm tr} [ \hat O(0) \hat U \mathcal{\hat A}_{\boldsymbol \alpha} \hat U^\dag]/2$, where we use the cyclic invariance under the trace. Thus, in order to calculate $\mathcal{O}_\alpha^W(t)$, we can evolve the many-body phase-point operator according to the von Neumann equation $d \mathcal{\hat A}_{\boldsymbol \alpha}/dt = -{\rm i} [\hat H, \mathcal{\hat A}_{\boldsymbol \alpha}]$. Making a product  ansatz for the phase-point operators, $\mathcal{\hat A}_{\boldsymbol \alpha} \approx \hat A^{[1]}_{\alpha_1} \otimes \hat A^{[2]}_{\alpha_2} \otimes \dots \otimes \hat A^{[M]}_{\alpha_M}$, and  assuming a general parametrization $\hat A^{[i]}_{\alpha_i} [{\bf r}_i(t)]=\wp [{\bf r}_i(t)]$ yields a coupled set of differential equations for ${\bf r}_i(t)\equiv (s^x_i,s^y_i, s^z_i)^T$.

For example, for the Ising interaction Hamiltonian
\begin{align}\label{H-Ising}
	H_{ZZ} =  \frac{1}{2} \sum_{n,m} J_{nm} \sigma^z_n \sigma^z_m
	,
\end{align} with $J_{nm}=J_{mn}$ and $J_{nn}=0$,  the classical (mean-field) equations for the spin components are then given by
\begin{align}
	\dot s^x_n &=   -  2s^y_n  \sum_ m J^z_{n,m} s^z_m \equiv -2s^y_n \beta^z_n \\
	\dot s^y_n &=  2 s^x_n  \sum_ m J^z_{n,m} s^z_m \equiv 2s^x_n \beta^z_n \\
	\dot s^z_n &= 0,
\end{align}
where we introduce the quantity  $\beta^{\alpha=x,y,z}_n\equiv \sum_ m J^z_{n,m} s^{\alpha=x,y,z}_m$ which can be interpreted as an effective magnetic field on spin $n$ induced by the mean-field interactions with the other spins. Solving  these equations yields:
\begin{align}\label{mf-s}
s_n^\pm(t) = s_n^\pm(0) \exp\Big(\pm 2i t \sum_j J_{nj} s^z_j\Big),
\end{align}
where $s^\pm=(s_n^x \pm i s_n^y)/2$.
For completeness, we also give the classical equations of motion that are used for the XY interaction in Eq.\eqref{H},
\begin{align}
	\dot s^x_n &=    s^z_n \sum_ m J^\perp_{n,m} s^y_m \equiv s^z_n \beta^y_n \\
	\dot s^y_n &=   - s^z_n  \sum_ m J^\perp_{n,m} s^x_m \equiv -s^z_n \beta^x_n \\
	\dot s^z_n &=  \sum_ m J^\perp_{n,m} (s^x_m s^y_n - s^y_m s^x_n) \equiv s^y_n  \beta^x_n - s^x_n \beta^y_n
	,
\end{align}
and for dynamics under a transverse field:
\begin{align}
	\dot s^y_n &=  -2 \Omega s^z_n \\
	\dot s^z_n &=  2 \Omega s^y_n.
\end{align}

\section{Correlation functions in the Ising model}
\label{app-corr}

The time evolution for the Ising Hamiltonian \eqref{H-Ising} can be solved exactly~\cite{Lowe-Norberg-1957,foss-feig_nonequilibrium_2013,worm_relaxation_2013}. For a local operator at site $k$ [$\sigma^\pm_k = (\sigma^x_k \pm i \sigma^y_k)/2$] the time evolution can be calculated as:
\begin{align}\label{timeev-sx}
\langle \sigma^\pm_k \rangle (t) = \frac{\langle \sigma^\pm_k\rangle(0)}{2^N}
\sum_{\substack{m_1 \ldots m_N \\ \in \{-1,+1\}} }
\exp \Bigl( \pm 2it \sum_{j=1}^N J_{kj} m_j \Bigr) .
\end{align}
Here each $m_i$ takes the values $-1$ and $+1$, and the sum runs over all $2^N$ possible combinations.
Comparing with Eq.~\eqref{mf-s} one sees that DTWA gives the exact time evolution in this case, when the sum is approximated via a random sampling of
$s^z$ taking the values $+1, -1$ (see discussion in the main text).

\begin{figure}[tb]
	\includegraphics[width=1.0\columnwidth]{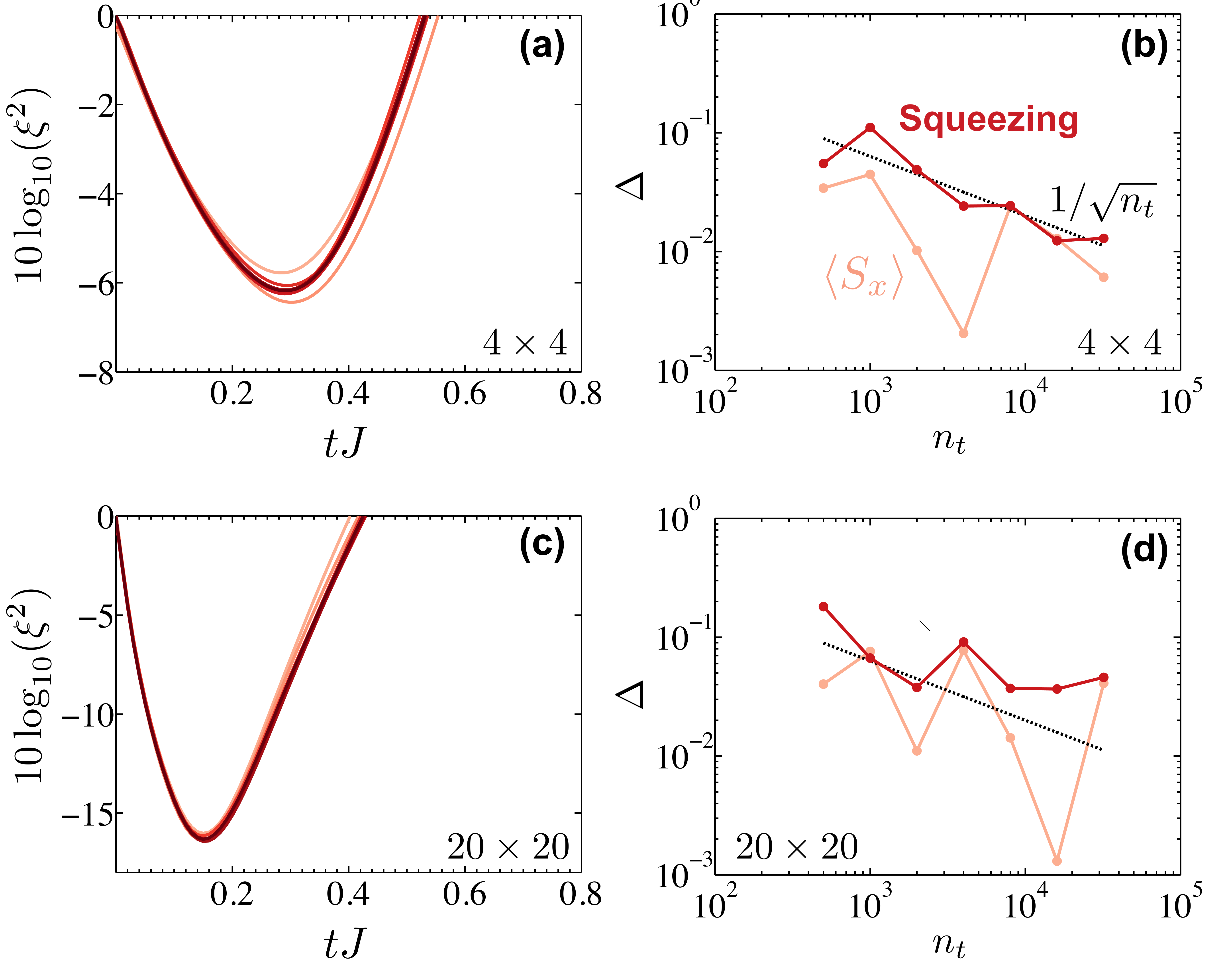}
	\caption{{\it Statistical convergence.} (a),(c) Time evolution of the spin squeezing parameter in DTWA for different number of trajectories: $n_t=500,$ $1000$, $2000$, $4000$, $8000$, $16000$, $32000$, $64000$ from light to dark. (a) Small $4\times 4$ system with XY interactions ($\alpha=1$), (c) $20\times 20$ system. In both cases for $n_t>4000$ curves are essentially indistinguishable. Panels (b),(d) show the maximum relative difference of the calculations from the $n_t=64000$ result. The upper line is for the spin squeezing parameter, the lower line for $\langle S_x\rangle$; there are smaller relative differences for this simple observable. Both differences decrease as $1/\sqrt{n_t}$ (dashed line) as expected. Panels (b),(d) are for the small and large system, respectively.}
	\label{fig:A1}
\end{figure}

The same calculation is possible for correlations. For example, between particle $i$ and $j$ ($i<j$):
\begin{multline}
\label{corr-exact}
{\langle \sigma^{\pm}_i \sigma^{\pm}_j \rangle(t)_{\rm exact}}=
 \frac{{\langle \sigma^{\pm}_i  \sigma^{\pm}_j \rangle (0) } }{2^{N-2}} \!\!
 \sum_{\substack{m_1 .. m_{i-1} m_{i+1} .. m_{j-1} m_{j+1} .. m_N \\ \in \{-1,+1\} }} \\
\exp \Bigl(\pm 2it \sum_{\substack{a=1\\a\neq i,j}}^N J_{ia} m_a \Bigr)
\exp \Bigl(\pm 2it \sum_{\substack{b=1\\b\neq i,j}}^N J_{jb} m_b \Bigr) ,
\end{multline}
Note that the two sums with $m_i$ and $m_j$ are missing. This is to be compared with the DTWA which estimates the same quantity (in the limit $n_t\to \infty$) as
\begin{multline}\label{corr-twa}
{\langle \sigma^{\pm}_i \sigma^{\pm}_j \rangle (t)_{\rm DTWA}}=
\frac{{\langle \sigma^{\pm}_i
\sigma^{\pm}_j \rangle (0) } }{2^{N}}
\sum_{s^z_1 \ldots s^z_N \in \{-1,+1\} } \\
\exp \Bigl(\pm 2it \sum_{a=1}^N J_{ia} s^z_a \Bigr)
\exp \Bigl(\pm 2it \sum_{b=1}^N J_{jb} s^z_b\Bigr).
\end{multline}
We note that Eqs.~\eqref{corr-exact} and \eqref{corr-twa} are valid for all combinations of signs $++$, $+-$, $ -+$, and $--$. One sees that the only difference between Eqs.~\eqref{corr-exact} and \eqref{corr-twa} is the  two additional sums with $s^z_i,s^z_j \in \{-1,+1\}$ in  Eq.~\eqref{corr-twa}.  This gives rise to an error of
\begin{align}
\label{eq:corr_error}
{
\langle \sigma_i^{\pm} \sigma_j^{\pm} \rangle (t)\bigr._{\text{DTWA}}
}
  =  \langle \sigma_i^{\pm} \sigma_j^{\pm} \rangle (t) \bigr._{\text{exact}} \, \cos^2 ( 2  t J_{ij}).
\end{align}
Since, for example, $\langle S_x^2 \rangle/N^2=\frac{1}{N} + \frac{1}{N^2} \sum_{i\neq j} \big( \langle \sigma_i ^+  \sigma_j ^+  \rangle + \langle \sigma_i ^+  \sigma_j ^-  \rangle + {\rm c.c.} \big)$, this correlation contains an error.  Analogously, it is straightforward to see that due to the conservation of the $z$ component of the spin there is no error made when calculating ${\rm Re} \langle S_z S_y \rangle = \sum_{i\neq j} 2 {\rm Re} \langle \sigma_j^z \sigma_i^-   \rangle$ with DTWA.

\section{Statistical convergence}
\label{app-conv}

The quantum noise in the DTWA calculations is introduced via a Monte Carlo sampling. For the success of the method in large systems it is important that the results converge when increasing the number of sample trajectories $n_t$.
We test, for example, the
time evolution of $S_x$ and the squeezing parameter
for different numbers of trajectories $n_t$,
for a small system with XY interactions ($\alpha=1$) and for a large system (see Fig.~\ref{fig:A1}).
In all cases for $n_t>4000$ the curves are essentially indistinguishable.
The maximum relative difference from the $n_t=64000$ result in these calculations
decreases as $1/\sqrt{n_t}$, as expected, and has the same magnitude for both the small and the large system. For the one-particle observable $S_x$, the convergence with increasing $n_t$ is faster than for the squeezing parameter, a two-particle observable.

\bibliography{twa}

\end{document}